# Estimates of neural health using cochlear implant digital twins correlate with speech recognition

Erin L. Bratu, Andrea J. DeFreese, Katelyn A. Berg, Margaret Wilson, Aaron C. Moberly, Robert F. Labadie, and Jack H. Noble, *Senior Member, IEEE*

***Abstract*—The cochlear implant is a neural prosthetic that uses an array of electrodes to directly stimulate auditory nerve fibers with electrical impulses. One factor that is known to impact outcomes is the quality of the electro-neural interface. In this work, we propose a pipeline that permits creating accurate and patient-specific digital twins that model the cochlear implant electro-neural interface. We leverage medical images and electrophysiological measurements from the patient's device for digital twin construction. We validate our digital twins by comparing predictions of neural response measurements to clinical measurements and speech recognition outcomes. Neural health estimated by the digital twin correlates strongly with word recognition rates (R=0.75, p<1e-5) in an analysis of 29 cases and sentence in noise recognition rates (R=0.78, p<1e-4) in an analysis of 22 cases. We also report preliminary findings with digital twin-based patient remapping with 6 subjects, where word recognition rates improved on average by 10 percentage points. These digital twins have the potential to provide clinicians with an unprecedented window into the electro-neural interface for individual cochlear implant recipients, which could lead to improved hearing outcomes when used as a foundation for program customization.**

## I. INTRODUCTION

THE Cochlear Implant (**CI**) is a neural prosthetic that uses an array of electrodes surgically inserted into the cochlea to directly stimulate auditory nerve fibers (**ANFs**) with electrical impulses [1], bypassing dysfunctions of the inner ear and restoring hearing sensation. The device consists of an external and internal component. The external portion consists of a processor, which controls how sounds detected by a built-in microphone are translated to instructions for the internal component. A transmitter is magnetically connected to an internal receiver/stimulator, which is placed between the scalp and skull during surgery. This component is connected to the remainder of the internal device, an array of electrodes inserted directly into the cochlea. These electrodes are activated according to the instructions received from the processor.

Despite numerous advancements and innovations in CI technology in the decades since its introduction leading to good outcomes for many recipients, individual outcomes remain highly variable [2-5]. Many CI recipients still receive little benefit from their device beyond basic sound awareness and lip-reading assistance.

One factor that is known to impact outcomes with CIs is the quality of the Electro-Neural Interface (**ENI**) [6-9]. To interface with the ANFs, CIs take advantage of the tonotopic nature of the cochlea, where ANFs located deeper (shallower) in the cochlea induce lower (higher) frequency pitch sound sensation. A band of sound frequencies is assigned to each electrode, and when sound energy is detected in that band, the corresponding electrode is activated. These frequency bands, as well as other parameters, can be modified by an audiologist to improve recipients' hearing. The other parameters include the set of electrodes available for stimulation and characteristics of the stimulus pulses delivered to each electrode, such as stimulus dynamic range, pulse width, and stimulation rate. Collectively, these settings are referred to as a patient's "map," which the processor uses to determine the instructions sent to the electrode array based on the frequency composition of a received sound.

Traditional clinical approaches to how audiologists create a

This paragraph of the first footnote will contain the date on which you submitted your paper for review, which is populated by IEEE. This work was supported by grant R01DC014037 from the National Institute for Deafness and Other Communication Disorders. The content is solely the responsibility of the authors and does not necessarily reflect the views of this institute.

Erin L. Bratu, Ph.D., is with the University of Rennes, Rennes, France (e-mail: erin.bratu@univ-rennes.fr).

Andrea J. Defreese, Au.D. (e-mail: andrea.defreese@vumc.org), Margaret Wilson (email: margaret.wilson.1@vumc.org), and Aaron C. Moberly, M.D. (e-mail: aaron.c.moberly@vumc.org) are with the Vanderbilt University Medical Center, Nashville, TN 37232 USA.

Katelyn A. Berg, Au.D., Ph.D. is with Washington University St. Louis, St. Louis, MO, USA (email: berg.k@wustl.edu).

Robert F. Labadie, M.D., Ph.D. is with the Medical University of South Carolina, Charleston, SC USA. (e-mail: labadie@musc.edu).

Jack H. Noble, Ph.D. is with Vanderbilt University, Nashville, TN 37235 USA. (e-mail: jack.noble@vanderbilt.edu).

recipient's map are primarily based on subjective feedback from a recipient, which can limit the ability to effectively and efficiently reprogram the implant and create a high-quality ENI. For many individuals, it requires weeks or months with a new map for the brain to adapt.

A comprehensive description of the unique ENI for each individual includes numerous variables that impact how stimuli created by the CI electrodes activate individual ANFs. These variables include the unique shape of the ear anatomy, the position of the CI within their cochlea, the electrical characteristics of the tissue, the health of the ANFs, and the stimulation settings in the recipient's map. Further, the ENI is time dependent within an individual recipient due to temporal physical dynamics (e.g., neural refractory behavior) where, following the activation of an ANF, it is difficult to reactivate the ANF for a period of 1-3ms following the initial activation [10, 11].

To create an ideal ENI would require the ability to control activation of individual ANFs with high selectivity, similarly to natural hearing. However, available CI devices leverage at most 22 contacts and far-field monopolar stimuli to activate the ANFs, which can number up to around 30,000 in a healthy ear. Compared to the high-degree of frequency selectivity afforded by natural hearing, reducing hearing stimuli to 22 or fewer channels creates substantial spectral resolution degradation. Further, the use of far-field stimuli results in broad and overlapping ranges of ANFs activated for each contact, which can lead to further spectral smearing artifacts, as well as channel interaction artifacts due to refractoriness[11, 12].

In this work, we present a workflow of algorithms to create digital twins of the ENI for individual CI recipients. We leverage computational models of the implanted cochlea that account for the unique anatomy, electrode placement, electrical characteristics, and ANF health of the individual. These digital twins have the potential to provide clinicians with an unprecedented window into the ENI for individual CI recipients, which could lead to improved hearing outcomes when used as a foundation for customized CI programming.

### *A. Previous Work*

Many computational models of the implanted cochlea have been developed over the course of several decades. For brevity, we will focus only on three-dimensional models that incorporate some degree of patient-specific features and the most common techniques used in their construction.

The first step in the creation of these models is creation of the custom spatial representation of the patient, starting with reconstruction of the anatomical structures in the inner ear. Many models utilize reconstructions created from high-resolution images, e.g., from photomicrography or micro-computed tomography (μCT), of cadaver cochleae as a foundation, which allows for more detailed structures than could be accurately obtained from imaging modalities that can be used *in vivo*. Then, the models are adapted to the lower-resolution *in vivo* patient images either rigidly [13-15] or elastically to account for non-rigid differences between individuals [16, 17]. Several groups have also proposed pipelines for localizing electrode positions from postoperative (usually CT) images [15, 18, 19].

To simulate the spread of electric potential in the inner ear, several numerical methods have been proposed to solve the volume conduction problem, e.g., using the finite element method [13, 16, 20], boundary element method [21-23], and finite difference method [24]. Most groups use a set of fixed tissue resistivity values, although it is possible to electrically customize the model at this step to account for patient specific tissue resistivities.

Finally, to simulate activation of individual ANFs in response to the electric field created by the CI, many groups rely on electric-circuit models of the ANF based on the "warmed" Hodgkin-Huxley model [25], such as the Rattay [26, 27], or the Frankenhaeuser-Huxley model, like the generalized Schwarz-Eikhof-Frijns (**gSEF**) model [22, 28, 29]. These models describe activation of single ANFs. As there can be up to 30,000 individual ANFs in the inner ear, simulating each one individually, even for relatively short stimulation sequences, quickly becomes impractical. Higher level phenomenological models have been proposed to synthesize ANF bundle-level behavior as the compound response to 100s or 1000s of ANFs [10]. Other groups have not explored patient-specific neural health representation at the ANF bundle level.

## II. METHODS

In this work, we propose a pipeline that permits accurate customization of the digital twin to individual patients, capturing patient-specific spatial, electrical, and neural health properties. We leverage not only medical images, but also electrophysiological measurement from the patient for digital twin construction. We validate our digital twins by comparing predictions of neural response measurements to clinical measurements and outcomes.

### *A. Data*

In this study, we create digital twins for 30 CI recipients. For each subject, we collected pre- and post-implantation CT. As a pre-processing step, ear anatomy was accurately localized in pre-implantation CT using existing validated methods[30-34]. Similarly, pre-existing methods were used for localizing CI electrodes in post-implantation CT and registering the pre- and post-implantation images to determine the precise intra-cochlear location of the contacts [18, 19, 35, 36]. We also collected several electrophysiological measures from the subject. First, "Stimulation Current Induced Non-Stimulating Electrode Voltage (**SCINSEV**)" measurements were collected. SCINSEVs are measurements of voltages present at all non-stimulating CI electrodes when a single electrode is activated as a current source [11].

Next, we collected neural response telemetry measurements, including Amplitude Growth Functions (**AGF**s) on every contact and Spread-of-Excitation (**SOE**) functions on even numbered electrodes. Neural response telemetry permits using the implant to measure electrically evoked compound action potentials (**eCAPs**) when using the CI electrodes to both stimulate and record the neural response. We rely on the forward-masking subtraction strategy for stimulation artifact reduction in neural response measurements [11]. With AGF, stimulation current is increased on a stimulating contact and the growth of the neural response is recorded. With SOE, one electrode at a time is chosen as the active stimulation electrode and a constant stimulus current is applied while other electrodes are chosen as masker for forward-masking subtraction. The resulting neural responses measure how much ANF stimulation overlap is shared between pairs of contacts.

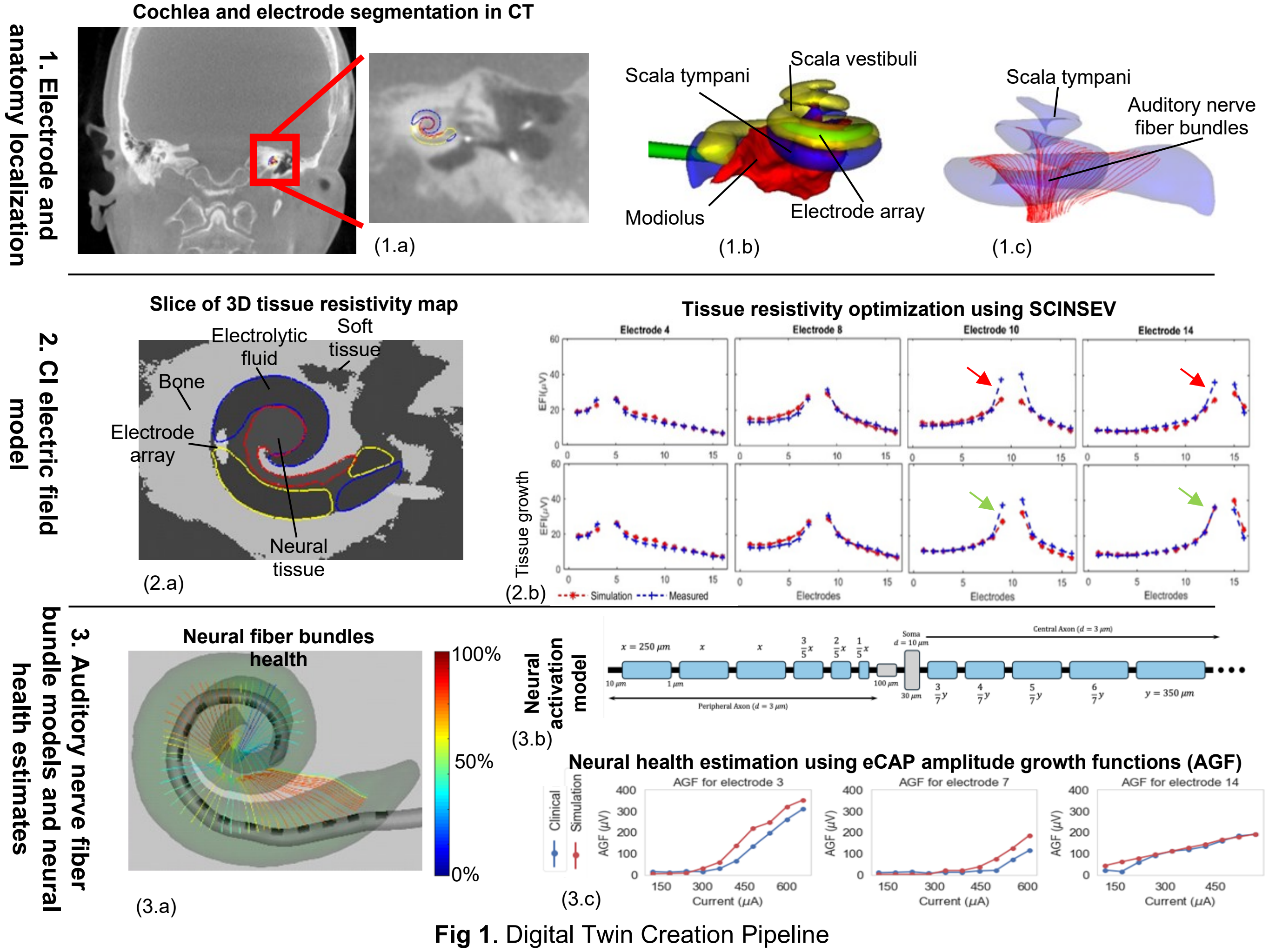

**Fig 1**. Digital Twin Creation Pipeline

AGF measures will be used for estimating patient-specific neural health and threshold parameters, while SOE measures will be held out as testing data to evaluate accuracy of model predictions.

## *B. Digital Twin Creation*

The digital twin creation pipeline we propose is summarized in Figure 1. As seen in the figure, the model is created in three steps: 1) spatial model, 2) electrical model, and 3) ANF model.

*Spatial model.* Creating a high resolution spatial model is challenging due to the small size of the cochlear features compared to the resolution of conventional computed tomography (**CT**) imaging, which is typically in the range of 0.2-0.4 mm. Although the external walls of the cochlea are visible in conventional CT, the boundaries of the internal structures are not resolved. To overcome the limitations of imaging methods that can be used in vivo, we use a statistical shape model (**SSM**) constructed from manual segmentations of sixteen µCTs of cadaver cochleae [31]. These µCTs have a voxel resolution of 36 µm isotropic, which is approximately ten times greater than that of conventional CT. In these µCT images, internal features of the cochlea are well-delineated, allowing for segmentations with much greater detail. The SSM is used to segment structures from a preoperative CT image, including the scala tympani, scala vestibuli, and modiolus, providing us with a high-resolution model of a patient's unique anatomy.

The second portion of the spatial model is the segmentation of the electrode array from postoperative CT. This process faces similar difficulties as segmentation of the anatomy, due to the small size of the electrodes and their close spacing. Additionally, metal artifacts from the array can distort the intensities of voxels surrounding the array, further complicating the process of identifying the electrode locations. Using information available from CI manufacturers about the morphology of the electrode array, our group has developed automatic methods for segmenting arrays with distantly-spaced and closely-spaced electrodes with sub-voxel accuracy, as described in [18, 19]. A spatial model including the cochlear anatomy and electrode array is shown in Figure 1.1.b.

The final component of the spatial model is the segmentation of auditory nerve fibers (ANFs). At approximately 2 µm in width, these fibers are incredibly small and cannot be identified visually, even with the higher resolution provided by µCT imaging. Instead, we use a technique based on *a priori* knowledge of ANF morphology to create these fibers. An initial, semiautomatic localization method for this task is described in [37]. More recently, we have introduced an improved automatic method that produces more realistic spiraling paths [38]. An example result of this process can be seen in Figure 1.1.c.

*Electrical model.* To obtain realistic estimates of voltage spread in the cochlea, we use a volume conduction model of the inner ear [17]. The foundation of this volume conduction model is the tissue class label map of the inner ear, a 3-D volume in which each voxel is assigned to a tissue class corresponding to the anatomical structure within which that voxel is found [17]. These label maps are created from the segmentations of cochlear structures produced by the spatial model (see Figure 1.2.a). To increase the resolution of these maps beyond that afforded by conventional CT, we make use of corresponding manual segmentations created from µCT images of nine cadaver cochleae. For each of these µCT specimens, we calculate thin-plate spline (TPS) transformations to register the segmented structures to those obtained from the patient imaging data. In addition to the manual segmentations, we have a tissue class label map for each cadaver specimen that contains labels for each structure as well as labels for bone, air, and the internal auditory canal (IAC). If we apply these transformations to these tissue class maps, we obtain nine tissue maps for our patient data. Majority voting produces the final map from those nine options with dimensions of $251 \times 251 \times 151$ and a voxel size of 72 µm isotropic. We subsequently add representations of the electrodes and their silicone carrier from the electrode positions localized for the spatial model and a mask of the carrier, respectively. In total, we represent five different tissue classes: electrolytic fluid, soft tissue, neural tissue, bone, and air. Voxels contained within the intra-cochlear cavities are assigned to electrolytic fluid, while voxels within the modiolus are assigned to neural tissue. The electrode array carrier is assigned as air, since silicone is nonconductive. The electrodes, which are made of metal, have a very low resistivity. However, due to the nature of the finite difference method we use to solve the volume conduction model, we also assign them to electrolytic fluid to prevent the potential introduction of numerical instability that can arise at the interface of materials with resistivity values that differ by several orders of magnitude. All remaining voxels not corresponding to bone, air, or electrodes are assigned to soft tissue. With this resistivity map, we can calculate the electric potential at the center of each voxel, henceforth referred to as nodes, by using the finite difference method (FDM) to solve Poisson's equation for electrostatics:

$$-\sigma\nabla^2\varphi = \nabla \cdot \mathbf{J}, \tag{1}$$

where $\sigma = \frac{1}{\rho}$ is the conductivity of each node, $\varphi$ is the electric potential, and $\boldsymbol{J}$ is the current density.

To customize our electrical model to each patient, we adapt the electrical resistivity values for electrolytic fluid and soft, neural, and bone tissues such that our simulated voltages reflect the patterns shown in clinical SCINSEV measurements. Our current approach utilizes a grid search for a defined parameter space around the default resistivity values. For each set of parameters in this grid, we simulate a voltage map for each of the electrodes in the array acting as the stimulus origin. We then sample each of these maps at the locations of the non-stimulating electrodes to create our simulated SCINSEV measurements. Once we have obtained the simulated SCINSEVs for each parameter combination, we use the root mean squared error (**RMSE**) between the simulated SCINSEV data and the clinical measurements to find the resistivity values that minimize this error. For a subset of subjects where SCINSEV is measured post-operatively, it is possible that scar tissue has accumulated around a portion of the array. This typically occurs on the basal portion of the array and is indicated by sharper increases in the voltages recorded on the electrodes adjacent to the stimulating contact (see red arrows in Fig. 1.2b). When we detect this, we add tissue growth to the model to improve the fit between the predicted and measured SCINSEV (see green arrows in Fig. 1.2b). The final step in creating our electrical model is to calculate voltage maps using these optimized parameters. Examples of clinical and simulated SCINSEVs using patient-customized resistivity parameters are shown in Figure 1.2.b.

*ANF model.* The final component of the digital twin is the neural response model, in which we simulate activation of ANFs. This section of the model can be further divided into two sub-components: the simulation of single-fiber APs and the simulation of fiber bundles that account for neural health. In the first sub-component, we use a biophysical ANF model to represent a single fiber at each of the 75 fiber bundles segmented as part of the spatial model. In this work, we use the generalized Schwarz-Eikhof-Frijns (**gSEF**) model [28, 29, 39] to simulate single fiber APs. This computational model implements a set of partial differential equations that describe ion channel kinetics in the nerve cell membrane and the propagation of a stimulus along the length of the fiber. These models take as input one or more time-varying sequences of voltages that represent extracellular stimulation, e.g., that from an intracochlear electrode. We create these sequences by sampling the 3-D voltage maps produced by the electrical model at a series of locations corresponding to the segmented ANFs. Solving the partial differential equations for these models with the specified input stimuli produces simulations of the cell membrane current along each fiber. We then scale these membrane currents by a transfer resistance that relates these locations to a recording electrode. These transfer resistances are obtained by solving the electrical model in much the same way as that used to obtain the initial voltage maps, with the only difference being that locations along the ANFs are used as the stimulus origins and the voltage maps are sampled at the electrodes. By recording membrane currents at each of $N$ locations along a single fiber at bundle $f$, we can obtain the single fiber action potential (SFAP) at a recording electrode $E$ due to stimulation from another electrode $S$ as $SFAP(f, S, E)$.

Next, we propose a statistical ANF bundle model which extends the SFAP predictions to model the behavior of a distribution of up to ~500 ANFs in each bundle. First, we use a binary search scheme to find the threshold stimulus current $T_{gSEF}(f, S)$ for each single fiber model for each fiber bundle $f$ and stimulating electrode $S$ as the minimum stimulus current that will result in a predicted SFAP intra-cellular voltage that dips below and returns above -20 mV. Because the gSEF model assumes a fixed fiber morphology, it is likely that patient-specific variations and variability in fibers along the length of the cochlea can impact thresholds. Thus, we rely on the clinically measured AGF to determine a per bundle scaling adjustment $s(f)$ to the threshold predictions. To determine $s(f)$, we first find the AGF threshold $AGF_t(S)$ for each stimulating electrode $S$ by fitting a linear function to the above noise floor eCAPs. $AGF_t(S)$ is then defined as the stimulation level where the linear function equals zero. We assume $1.5AGF_t(S)$ should correspond approximately to the level that would stimulate half the fibers in the easiest to excite bundle for that contact. To enforce $s(f)$ to be smoothly varying, we define 4 Gaussian radial basis functions (RBFs), uniformly spaced over the 75 bundles, as control points. Next, we find $\vec{p}$, the vector of 4 RBF magnitudes, as the least-squares solution for $\text{argmin}_{\vec{p}} \left( (RBF_t\vec{p})(S) - 1.5AGF_t(S)/T_{gSEF}(f_{\min}, S) \right)^2$, constrained to a maximum of 2.5. Then, we define $s(f) = RBF_t\vec{p}$, and we determine the final threshold map for the patient as $T(f, S) = s(f)\, T_{gSEF}(f, S)$. $T(f, S)$ maintains the shape and structure predicted by the gSEF model but the results for each bundle are scaled equally across electrodes to better match clinically measured, patient-specific threshold behavior.

Next, inspired by [10], we define the proportion of ANFs in a bundle that are activated in response to a given stimulus current $I$ using the cumulative distribution function as

$$BP_f(S, I) = \frac{1}{\sigma\sqrt{2\pi}} \int_{-\infty}^{I} e^{\frac{-(x - T(f,S))^2}{2\sigma^2}} dx, \quad (2)$$

where $\sigma$ is assumed to be $T(f, S)/4$. This distribution function represents a simplified model of variability in stimulation responsiveness of ANFs within a bundle due to both physiological variability as well as stochasticity. Using this bundle activation proportion function, we can estimate the bundle action potential amplitude as the product,

$$BAPA(f, S, E, I) = N(f)\; N1P2(f, S, E)\; BP_f(S, I), \quad (3)$$

where $N(f)$ is the estimate of how many ANFs in bundle $f$ are available for activation (discussed further below), and $N1P2(f, S, E)$ is the amplitude of the estimated SFAP at bundle $f$ for stimulating electrode $S$ measured by recording electrode $E$. Finally, the compound action potential amplitude is estimated as the sum of the bundle action potential amplitudes over all bundles:

$$CAPA(S, E, I) = \sum_f BAPA(f, S, E, I) \quad (4)$$

$N(f)$, the number of ANFs in a bundle that are available for activation, can be impacted not only by ANF health, but also by temporal neural properties, such as refractoriness, accommodation, and facilitation. These properties do not need to be modeled in order to capture the behavior of the nerves for AGF or SOE neural response telemetry measurements as these measurements are conducted from a neural steady state. Thus, while important, we leave modeling of temporal behavior to future work. For the scope of this work, we define the number of ANFs available for stimulation in a bundle $f$ in steady state as $N(f) = H_f$, where $H_f$ is the estimate of the total number of healthy ANFs within bundle $f$.

To customize the ANF model to individual CI recipients, we optimize $H_f$ with the goal of minimizing error between model predicted and clinically measured AGF as shown in Figure 1.3.c. AGF can be used for this because AGF slopes are known to be sensitive to neural health [11].

The objective function we propose for optimization is based on minimizing mean squared error between model predicted and clinically measured AGF with some modification for solution regularization and greater emphasis of fit for datapoints that are above the noise floor. We consider eCAPs with magnitude less than 20 µV as falling below the noise floor. The overall objective function for the optimization is:

$$C(\overrightarrow{p_h}) = C_{AGF}(\overrightarrow{p_h}) + \alpha C_h(\overrightarrow{p_h}), \quad (5)$$

where

$$C_{AGF}(\overrightarrow{p_h}) = \frac{1}{N} \sum \omega \left( CAPA(S, E, I) - AGF(S, E, I) \right)^2, \quad (6)$$

is a mean of weighted squared errors over all $N$ AGF eCAP measurements available for the subject, including all stimulation current values $I$ for each stimulation and recording electrode combination, $(S,E)$. The weight $\omega$ is chosen to be 1 if the measured eCAP is above the noise floor and 0.4 otherwise.

$\overrightarrow{p_h}$ are the health parameters. Instead of defining independent parameters for each of our 75 fiber bundles, we define limited degrees of freedom in the parameters for regularization. We define 12 RBFs uniformly spaced over the 75 bundles as control points. Thus, our optimization problem has 12 total parameters. For regularization of parameters, we use

$$C_h(\overrightarrow{p_h}) = \sum_j \nabla p_{h,j}, \quad (7)$$

where the gradient operator is implemented with finite differences, to reward smoothly varying neural health parameters. The nerve bundle health is estimated as $H_f = RBF_h\overrightarrow{p_h}$. In equation 5, weighting parameter $\alpha$ is fixed at 1e-3. This parameter was determined heuristically to balance regularization and fit to the data.

We optimize $\overrightarrow{p_h}$ using an L-BFGS optimization routine in pytorch [40]. Equation 5 is implemented using differentiable operations so that auto-grad can be used to determine gradient descent directions. We use a learning rate of 5e-3, minimum change tolerance of 1e-6, and run the optimization for a maximum of 2500 iterations.

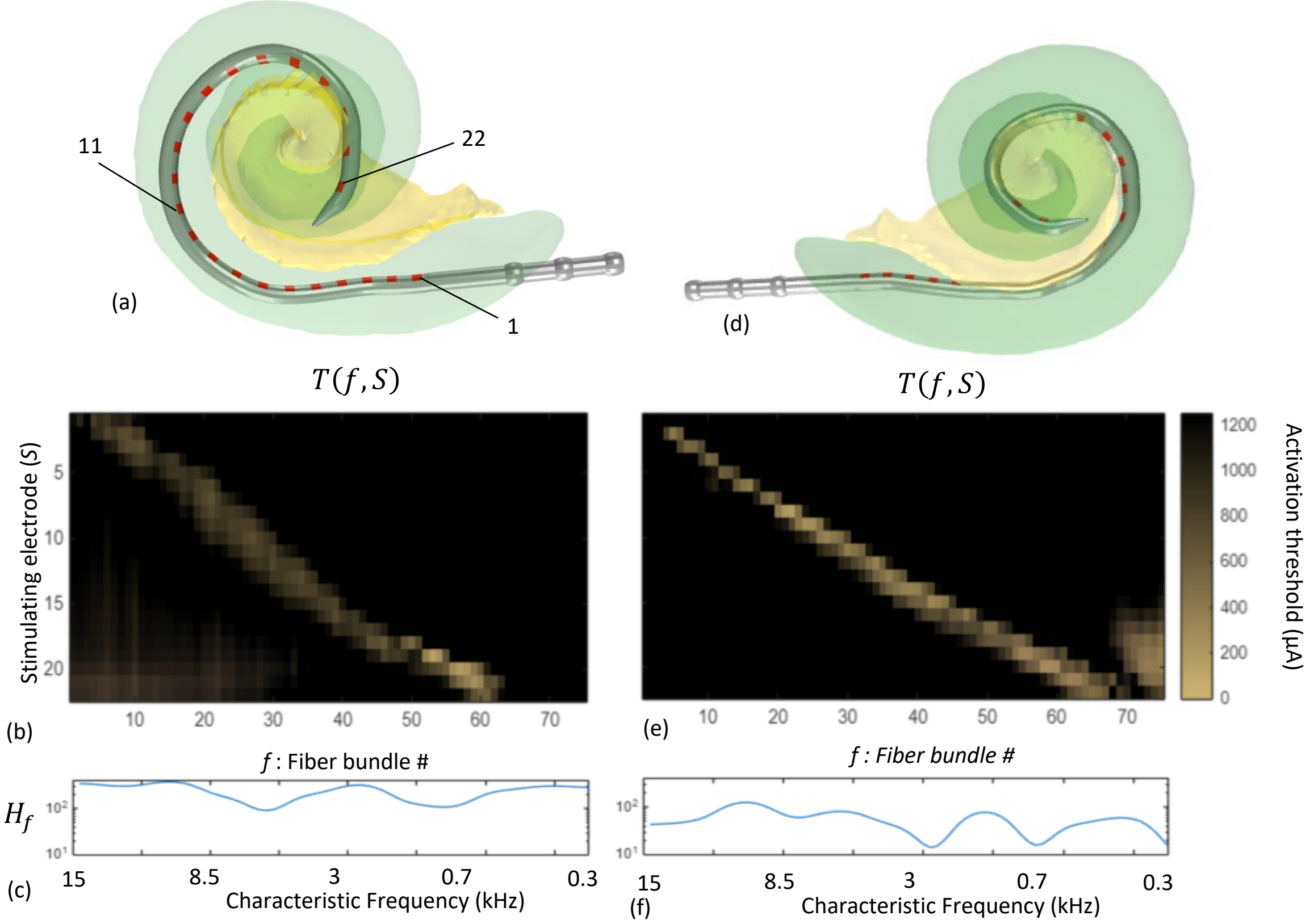

Figure 2. An example electro-neural interface result described by $T(f,S)$ and $H_f$. In (a), electrode position (red) is shown inside the scala tympani (green surface) relative to the modiolus (yellow surface), which houses the ANFs. In (b), $T(f,S)$ is shown, where the vertical axis corresponds to the stimulating electrode and the colors correspond to activation thresholds. In (c), the vertical axis corresponds to the count of healthy ANFs per bundle. In both (b) and (c), the horizontal axis corresponds to fiber bundle number, which is also mapped to ANF characteristic frequency. Panels (d)-(f) show similar information to (a)-(c) for a second example subject.

### C. Experiments

Our dataset includes 30 CI recipients with the required data to create a digital twin (post-op CT, SCINSEV, and AGF). The study protocol was performed under Vanderilt IRB approval 150417, and informed consent was obtained from all individual participants included in the study. Among the 30 study participants, 29 also had eCAP SOE measurements.

In our first experiment, the digital twin is used to predict SOE and these predictions are compared to measured SOE data to validate the ability of the model to predict eCAP measures on which it was not trained.

In our 2nd experiment, we investigate how well the digital twin estimates of patient-specific neural health correlate with speech recognition, in both quiet and noisy listening conditions. We propose an overall neural health score, *NHS*, for each subject, as:

$$NHS = \frac{1}{75}\sum_{f=1}^{75} \log\left(\frac{H_f}{\gamma} + \varepsilon\right), \quad (9)$$

where $H_f$ is the number of fibers estimated to be healthy in fiber bundle $f$, $\varepsilon$ is a small number to prevent instability when $H_f$ is zero, and $\gamma$ represents the number of fibers required for a healthy bundle. We propose using the log function to make *NHS* more sensitive to having small poor neural health regions than a simple average would be. In all our experiments we set $\gamma = 50$. All subjects included in this analysis had at least 6 months use of their device, since most CI recipients' performance plateaus after 6 months of experience [7].

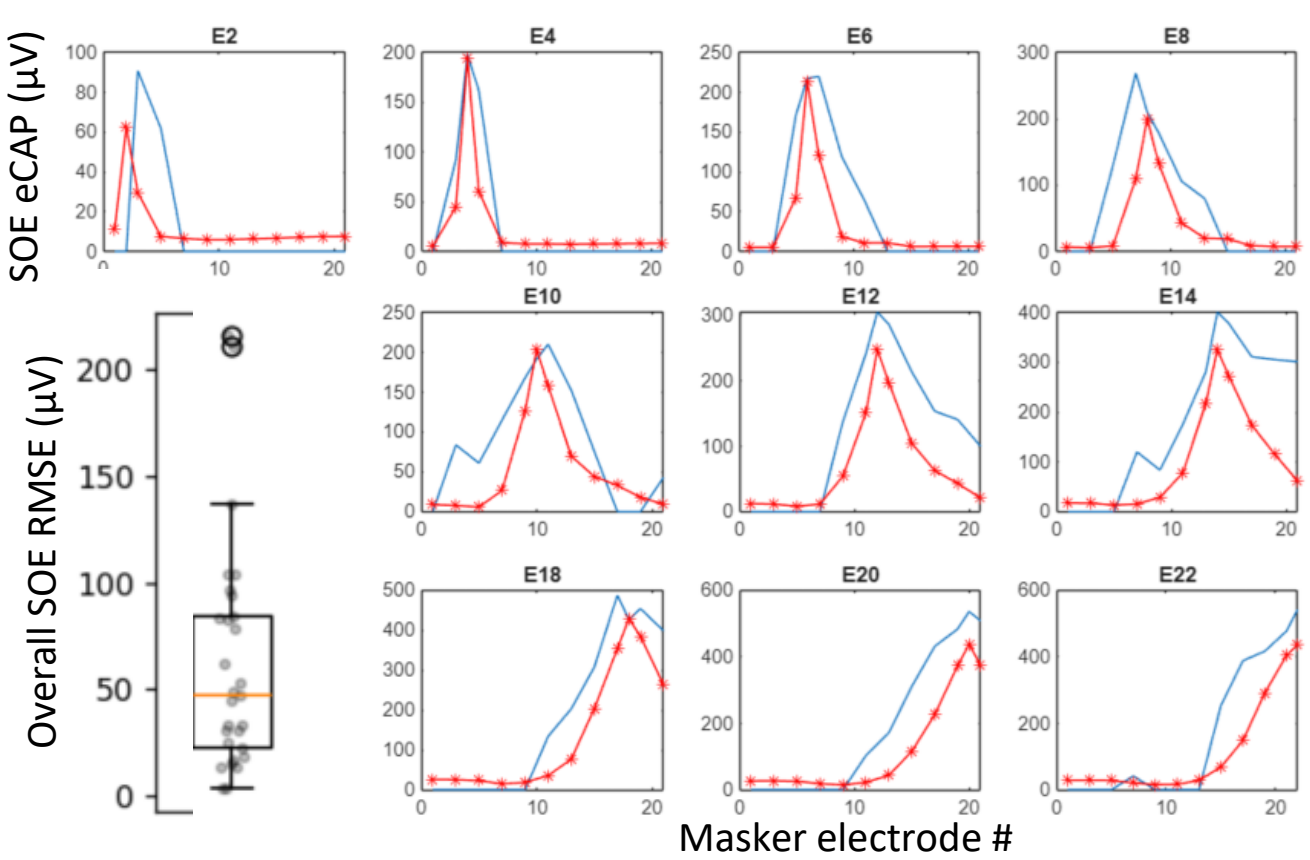

Figure 3. The lower left boxplot shows the overall RMS errors in the digital twin predicted SOE eCAP amplitudes compared to clinical measurements across all 29 included cases. The remaining plots show the digital twin predicted (red) and clinical (blue) SOE curves for the single subject with the worst RMSE (215 μV). Each plot is labelled according to the probe electrode, where the masker varies as shown on the horizontal axis.

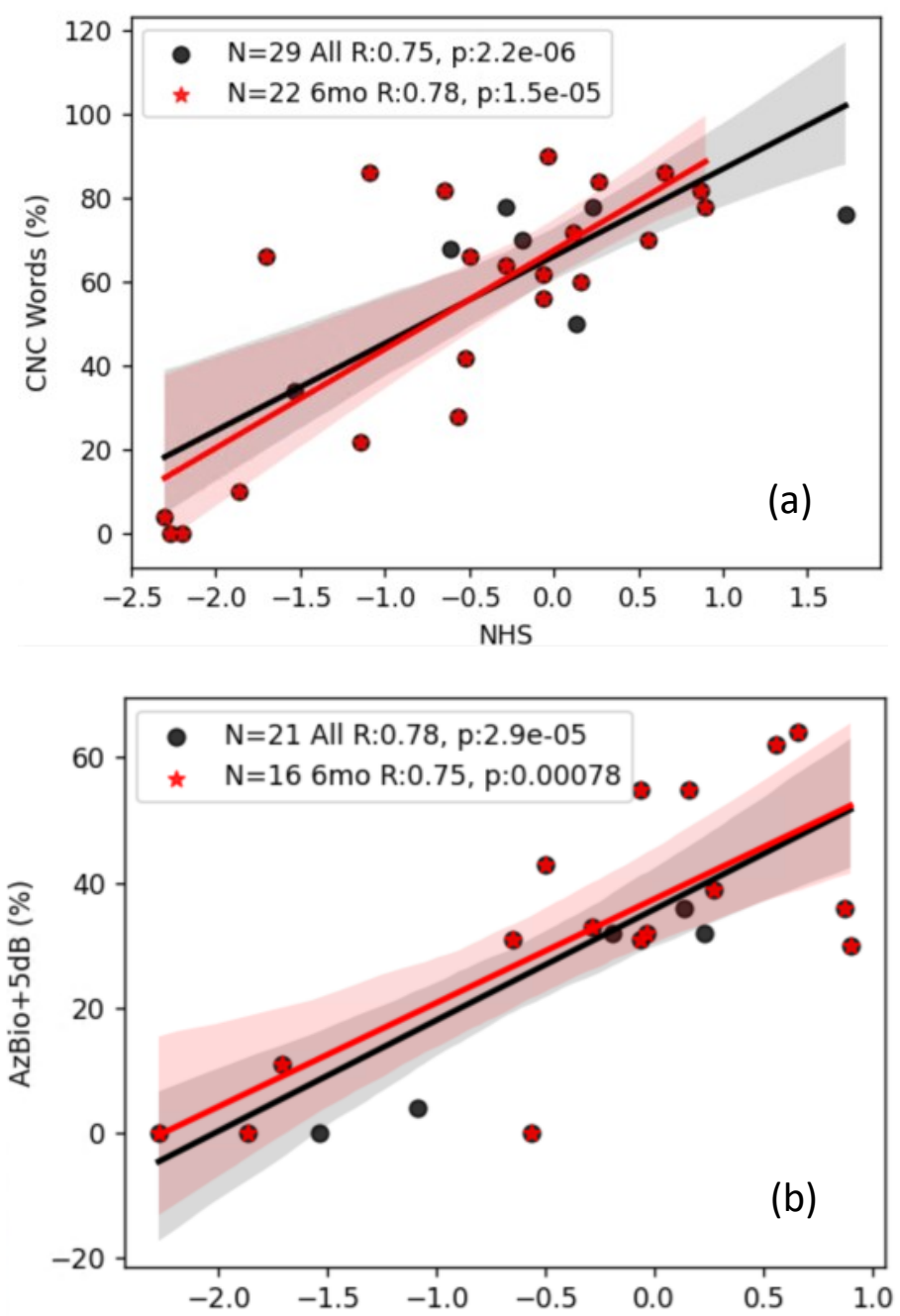

Figure 4. Scatter plots showing correlation between the digital twin neural health score (NHS) and speech recognition using clinical map: (a) CNC Word recognition in quiet; (b) AzBio Sentence recognition in +5dB SNR using multi-talker babble noise. Shown in red are measures taken at the 6 month post-surgery appointment. In black, we add to this the most recently available measures for subjects where 6 month data were missing.

In our final experiment, we show preliminary findings with prospective patient remapping based on the neural health estimates. We hypothesize that unhealthy regions experience more substantial channel interaction effects. With 6 participants, we identified the region where neural health is predicted to be lowest and deactivated half of the contacts in these regions in a uniform pattern. We then compare speech recognition scores measured at one or more following timepoints with clinical baseline scores.

## III. Results

In Figure 2, we show a $T(f,S)$ and $H_f$ pair from two example cases. These two arrays form the basis of the representation of the ENI in our digital twin model. As seen in the figure, $H_f$ contains an estimate of how many healthy fibers exist in each bundle. $T(f,S)$ describes, for each stimulating electrode $S$, the stimulation threshold level at which ANFs in each bundle are likely to be activated. In the color-mapped visualization of the matrix, it becomes straightforward to see where adjacent electrodes are likely to stimulate overlapping fiber bundle populations.

In experiment 1, we observe accurate model predictions of SOE eCAP measurements, with median RMSE below 50 µV. Even in the worst case with RMSE=215 µV (shown in Figure 3), reasonably good agreement in the overall shape and magnitude of the SOE curves can be observed.

In experiment 2, we observe significant correlation between digital twin derived neural health scores (*NHS*) and speech recognition rates achieved with clinical mapping. For 22 subjects, Consonant-Nucleus-Consonant (**CNC**) word recognition rates measured 6-months post implantation were available [41]. These data are shown in red in Figure 4(a). As seen in the figure, a significant correlation coefficient of R=0.78 is observed for this group. For another 7 subjects, these measures were available for a later time point, post-implantation. Shown in black in the same figure is the dataset supplemented with these additional datapoints. For 16 subjects, AzBio sentence recognition rates in +5dB signal-to-noise-ratio using multi-talker babble measured at 6-months post-implantation were available [42]. These data are shown in red in Figure 4(b). A similar significant correlation to that seen for the CNC word scores (R=0.75) can be observed. For an additional 5 subjects, these measures were available from a later time point. Shown in black is the dataset supplemented with these additional datapoints.

In experiment 3, for each subject, between 2 and 6 electrodes were selected for deactivation (out of 22). The resulting progression of word recognition scores, as measured using the CNC word recognition test [41], are shown in Figure 5. All subjects started with at least 1 year of experience with their clinically programmed CI and had achieved stable performance. Due to the preliminary nature of this initial data collection, follow-up measurements were not collected at consistent timepoints across subjects. Overall, uniform improvements in CNC word recognition can be observed in 4 of 6 cases. Of the remaining 2, in 1 case, a moderate decline is observed acutely followed by an overall increase after 1 day of experience. In the other case, a moderate increase is observed acutely followed by decline back to baseline after 1 day of experience. In all cases, no substantial decline in performance is observed, and all subjects reported preferring the new map and elected to keep it long term. The black line in the figure corresponds to the subject whose ENI is depicted in Figure 2a-c. For this subject, odd numbered electrodes from E5-E15, which span the deepest valley in $H_f$ seen in Fig 2.c, were deactivated. The cyan line corresponds to the subject whose ENI is depicted in Figure 2d-f. For this subject, E14, E16, and E19 were deactivated, corresponding to the two valleys in $H_f$ seen in Fig 2.c near bundles 40 and 55.

## IV. Conclusions

In this work, we propose a pipeline that permits creating accurate and patient-specific digital twins that model the cochlear implant electro-neural interface. The digital twins model the patient-specific spatial, electrical, and neural health properties. We leverage not only medical images, but also electrophysiological measurements from the patient's device for digital twin construction. Direct validation of the digital twins is not possible without post-mortem histological analysis. Instead, we indirectly validate our digital twins by comparing predictions of neural response measurements to clinical measurements and speech recognition outcomes.

In experiment 1, we found good agreement between clinically acquired SOE eCAP measurements and those predicted by the model. This suggests our digital twins are well capturing overlapping stimulation patterns by multiple electrodes.

In experiment 2, we found that digital twin estimates of neural health have significant correlation with speech recognition. The neural health score explains 50-60% of the variability in speech recognition ability in both quiet and noisy conditions in this dataset. This finding is significant for the cochlear implant field as finding factors that explain a significant portion of variability in outcomes has been a focus of substantial research effort for decades [7, 43-48]. A single factor explaining more than 50% of the variability in speech recognition is the strongest correlating factor reported in the literature, to the best of our knowledge. In [48], a similar level of correlation is reported between speech reception thresholds and eCAP inter-phase-gap slope effect. However, this was after up-weighting data from subjects with stronger eCAPs and paring down the dataset to 9 ears of 6 subjects from 24 ears of 13 subjects after removing subjects with lower eCAP responses. In our study, we did not exclude subjects with lower eCAP responses. Electrocochleography as a measure for cochlear hair cell health has also been shown to capture just under 50% of the variability in word recognition rates [49]. It is noteworthy that two independent measures of cochlear health are found to account for speech recognition at similar rates in separate studies. Future work will be aimed at combining *NHS* with other factors known to explain outcome variability in a multi-variable model.

Substantial work in the field has also been focused on using eCAP measurements to interrogate neural health and the ENI [50-52]. Reliable methods have remained elusive as eCAPs are a function of not only neural health but electrode position and tissue resistivity. In this work, we develop a comprehensive physical model that accounts for the spatial and tissue resistivity contributions to eCAPs, allowing for neural health to explain the remaining variability in AGF functions. Thus, it provides a physically plausible explanation for the observed AGF measurements. This could be especially beneficial for programming in the pediatric population or for any CI recipients where providing subjective feedback to the audiologist is difficult, as our digital twins are constructed entirely from objective measurements. However, one limiting factor is that we rely on post-operative CT imaging to determine electrode position, and this requires a certain amount of ionizing radiation.

In experiment 3, promising results are observed in a preliminary prospective remapping study, with 5 of 6 cases ending the study with higher CNC word recognition scores compared to their clinical baseline, and with the remaining case performing relatively well initially and ending the study with no change. Overall, CNC scores were improved on average by 10 percentage points. These results motivate conducting a formal controlled prospective clinical trial on neural health-informed remapping, which is being planned as future work.

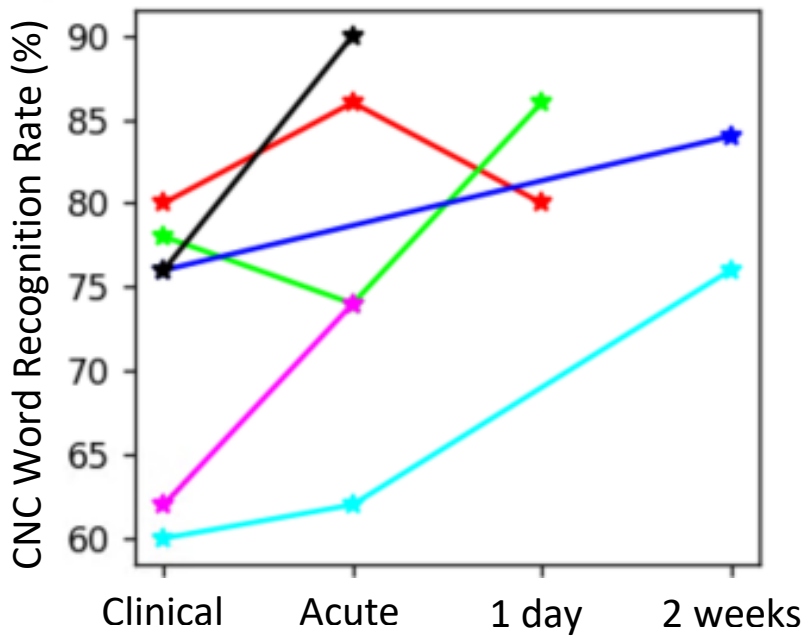

Figure 5. CNC word recognition rate progression for prospective remapping study. Each of 6 subjects is assigned a unique color.

There are several limitations to the current work. Our population model (Eqns. 3 and 4) for estimating compound action potential amplitude assumes a simplified additive model where N1P2 amplitudes add together synchronously. In reality, fiber populations can activate with different timing and without perfect synchronization. Due to this limitation, it is likely our estimates of the number of healthy fibers in each bundle, $H_f$, is an underestimation. Another limitation is that eCAP modeling was limited to steady state neural activation behavior in this work. Ongoing work is aimed at accounting for temporal artifacts in the model, including mixed synchrony, facilitation, accommodation, and refractoriness. These models will permit assessing the quality of, and optimizing, stimuli in a patient-specific manner. Another limitation is that we only analyze data from implants from a single manufacturer. Repeated experiments with other devices will be done in future work.